\documentclass[journal]{IEEEtran}
\usepackage{times,amsmath,epsfig,latexsym,amssymb,psfrag}
\usepackage{booktabs}
\usepackage{citesort}
\usepackage{subcaption}
\usepackage{paralist}
\usepackage[ruled,vlined]{algorithm2e}
\usepackage{float}
\restylefloat{table}

\usepackage{float}
\floatstyle{plaintop}
\restylefloat{table}

  \usepackage{setspace}

\newenvironment{definition}[1][Definition]{\begin{trivlist}
\item[\hskip \labelsep {\bfseries #1}]}{\end{trivlist}}

\newcommand{\qed}{\nobreak \ifvmode \relax \else
\ifdim\lastskip<1.5em \hskip-\lastskip
\hskip1.5em plus0em minus0.5em \fi \nobreak
\vrule height0.75em width0.5em depth0.25em\fi}

\begin{document}

\title{Network Lifetime Maximization for Cellular-Based M2M Networks}
\author{ Amin Azari and Guowang Miao\\
KTH Royal Institute of Technology\\
Email: \{aazari, guowang\}@kth.se}
\maketitle


\begin{abstract}
High energy efficiency is critical for enabling massive machine-type communications (MTC) over cellular networks.  This work is devoted to energy consumption modeling, battery lifetime analysis,  lifetime-aware scheduling and transmit power control for massive MTC  over cellular networks. We consider a realistic energy consumption model for MTC and model network battery-lifetime. Analytic expressions are derived to demonstrate the impact of scheduling on both the individual and network battery lifetimes. The derived expressions are subsequently employed in the uplink scheduling and transmit power control for mixed-priority MTC traffic in order to maximize the network lifetime. Besides the main solutions, low-complexity  solutions with limited feedback requirement are investigated, and the results are extended to existing LTE networks. Also, the energy efficiency, spectral efficiency, and network lifetime tradeoffs in resource provisioning and scheduling for  MTC over cellular networks are investigated.  The simulation results show that the proposed solutions can provide substantial  network lifetime improvement and network maintenance cost reduction in comparison with the existing scheduling schemes. 

\end{abstract}

\IEEEpeerreviewmaketitle


\section{Introduction}

\IEEEPARstart{I}{nternet} of Things (IoT) refers to the ever-growing network of uniquely identifiable smart physical objects that are capable of sensing or acting on their environment. 
Cellular IoT, IoT embedded in cellular network infrastructure, is expected to play a critical role in the success of IoT because cellular networks provide ubiquitous  coverage and roaming \cite{w_sony}. 
Cellular machine-to-machine (M2M) communications, also known as machine-type communications (MTC), means the communications of machine devices in cellular networks without human intervention and serves as the foundation of cellular IoT. The continuing growth in demand from cellular-based M2M 
communications encourages mobile network operators to investigate evolutionary and revolutionary radio access technologies for accommodating M2M traffic 
in cellular networks \cite{w_sony}. M2M communications are generally characterized by the massive number of concurrent active devices, small  payload size, and vastly diverse quality-of-service (QoS) requirements \cite{sysreq}. Moreover, in many M2M applications smart devices are battery driven and once deployed, their batteries will never be replaced. Then, long battery lifetime is crucial for them, especially when deployed in remote areas. Based on the 5G envision by Nokia \cite{nok}, bit-per-joule energy efficiency for machine-type communications must be improved by a factor of ten in order to provide battery lifetimes around 10 years.

\subsection{Literature Study}
\subsubsection{MTC over Cellular Networks}
 Random access channel (RACH) of the LTE-Advanced (LTE-A)  is a typical way for machine nodes to directly access the base station (BS). 
The capacity limit of RACH for serving M2M communications  is investigated in \cite{laya}, and it is shown that RACH is neither a scalable nor an energy-efficient access scheme for massive M2M communications. Clustered-access is investigated  in \cite{tcom} to reduce congestion in an overloaded condition. In \cite{zo2}, access class barring with multiple transmit power levels is introduced in order to reduce congestion using the capture effect at the BS. In \cite{zo1}, energy efficient random access for machine nodes in a multi-cell scenario is investigated, where the choice of serving BS and transmit power level are to be optimized. When a device successfully passes the RACH,  it can send scheduling request to the BS through the physical uplink control channel (PUCCH). Then, the BS performs the scheduling and sends back
the scheduling grants through the corresponding physical downlink
control channel (PDCCH). Now, the granted machine node is able to send data over the granted physical uplink shared channel (PUSCH). This scheduling procedure performs well in existing cellular networks for a limited number of long communications sessions, such as voice and web streaming. However, regarding the fundamental differences in  characteristics and QoS requirements of M2M communications, it is evident that the presented scheduling procedure cannot survive with a massive number of short-lived M2M communications sessions. 
The 3GPP LTE has defined some research projects to support low-cost massive machine-type communications in cellular networks. The development of LTE for low-cost massive MTC has been initiated in release 12,  and will be continued in release 13 \cite{nok1}. The target for LTE release 13 includes LTE category M (LTE-M), and narrow-band LTE-M (NB LTE-M) deployments, which are expected to offer MTC over 1.4 MHz and 200 KHz   bandwidths \cite{nok1}. 

\subsubsection{MTC Scheduling over Cellular Networks}
Scheduling is the process performed by the BS to assign radio resources to UEs. In general, scheduling is not part of the standardization work,  and is left for vendor implementation. However, signaling is standardized, and hence, any scheduling scheme should comply with the control requirements in the standards. 
Regarding the limited capacity of PDCCH, the number of UEs that can be served at once are limited. Then, the scheduling problem can be broken into two subproblems: (i) time domain scheduling, in which a subset of devices is chosen to be scheduled; and (ii) frequency domain scheduling, in which the available resource elements are allocated to the selected subset of UEs. 
A thorough survey on LTE scheduling algorithms for M2M traffic is presented in \cite{cusm}. This survey indicates that existing scheduling algorithms could be categorized into 4 main categories with regard to the scheduling metric as follows \cite{cusm}:
(i) channel-based schedulers, in which UEs with the highest signal to noise ratio (SNR) have priority in resource allocation in order to minimize the bit error rate and maximize the system throughput \cite{chan}; (ii) delay-based schedulers, in which  the delay budget  prioritize devices for resource allocation \cite{del,mostafa}; (iii) fairness-based schedulers, which are designed to guarantee a  fair distribution of radio resources among UEs \cite{fair}; and (iv) hybrid schedulers, which consider a combination of the aforementioned metrics as well as other metrics like power consumption \cite{coex}, buffer status, and data arrival rates \cite{cusm}. 


\subsubsection{Energy-Efficient MTC Scheduling}
While providing scalable yet energy efficient communications is considered as the key requirement for successful deployment of MTC over existing cellular networks \cite{nok,laya}, a limited number of research works has been focused on energy efficient uplink MTC scheduling. 
Energy efficiency of M2M communications over LTE networks is investigated in \cite{ltewang}, and it is shown that LTE
physical layer is not optimized for small data communications. Power-efficient uplink scheduling for delay-sensitive traffic over LTE systems is  investigated in \cite{lco}, where the considered traffic and delay models are not consistent with the MTC characteristics \cite{sysreq}, and hence, the derived results cannot be used here. Power-optimized resource allocation for time, frequency, and code division multiple access (TDMA, FDMA, CDMA) systems  has been investigated in \cite{poshti}. 
Uplink scheduling for LTE networks with M2M traffic is investigated in \cite{coex}, where the ratio between the sum data rates  and the power consumptions of all users is  maximized. In \cite{coex}, the authors have considered a simple model for energy consumption considering only the transmit power for reliable data transmission and neglected the other energy consumptions by the operation of electronic circuits which are comparable or more dominant than  the energy consumption for reliable data transmission \cite{guo}. In \cite{scalam}, a clean slate solution for dense machine deployment scenarios is proposed in which, each communications frame is divided into two subframes. The first subframe is dedicated to  the contention of machine nodes for access reservation, and the later is dedicated to scheduled data transmission of successful nodes using TDMA scheme. 
To the best of our knowledge, accurate modeling of energy consumption in machine-type communications, individual and network battery lifetime models, and corresponding scheduling algorithms are absent in literature.  As an extension of \cite{tcom}, which investigates clustered-access for massive M2M, in \cite{adhoc} joint energy efficient clustering and scheduling has been investigated, i.e. the cluster-size, selection of cluster-heads, and the amount of scheduled resources to cluster-heads have been optimized to prolong the battery lifetime.  In \cite{wcnc,vtc}, preliminary studies on feasibility of battery lifetime-aware scheduling  for  unclustered  M2M communications have been presented, and two exhaustive search algorithms for scheduling over frequency domain resources have been developed.  Substantial extension to \cite{vtc,wcnc} has been made in this paper, where sophisticated  scheduling algorithms  over {time/frequency}  resources along
 with  {low-complexity}  and  {limited feedback} solutions  have been developed under {different network lifetime definitions}. Furthermore, detailed analytical analysis, derivation of closed-form scheduling expressions, and complexity and fairness analysis have been presented in this paper.

\subsection{ Contributions }
 The main contributions of this paper include:

\begin{itemize}

\item
 Introduce accurate  energy consumption, and individual and network lifetime models for machine-type devices deployed in cellular networks by taking both transmission and circuit energy consumptions into account. 

\item
Present a battery lifetime aware resource allocation framework. 
Explore  MTC scheduling based on the Max-Min lifetime-fairness, and analyze its contribution in reducing the maintenance costs of   M2M networks.

\item
Present  uplink scheduling solutions for  MTC over single-carrier frequency division multiple access (SC-FDMA)  systems. Present {\it low-complexity}  scheduling  solutions with limited feedback requirement.  
\item
Figure out the energy efficiency, spectral efficiency, and network lifetime tradeoffs in uplink MTC resource provisioning and scheduling.

\item
Extend the proposed  solutions for existing 3GPP  LTE networks. Present lifetime-improvement evidence using simulation results in the context of LTE.




\end{itemize}

The rest of this paper is organized as follows. In the next section, the system model is presented. The battery lifetime-aware scheduling framework, and the coupling between network lifetime and  control parameters are presented in section III by investigating M2M scheduling  in time domain for narrow band cellular M2M networks.  The general scheduling problem with time/frequency domain radio resources is investigated in section IV. Low complexity scheduling solutions with limited feedback requirement are investigated in section V. As an example of lifetime-aware scheduling, in section VI we apply the derived solutions in section IV-V to the 3GPP LTE networks, and provide simulation results in section VII in order to demonstrate the lifetime improvement. Concluding remarks are given in section VIII.

\section {System Model}\label{sys}
 Consider a single cell with one base station and a massive number of  machine nodes, which are uniformly distributed in the cell. 
The machine nodes are battery driven and once deployed, their batteries won't be replaced, then long battery-lifetime is crucial for them. Consider the  uplink scheduling problem at time $t$, where a set of devices, denoted by $\mathcal A$, is to be served using a limited set of resources.  As we aim at deriving network lifetime maximizing solutions, both individual and network battery-lifetime metrics are defined.

\subsection{Lifetime Metric}
For most reporting MTC applications, the packet generation at each device can be modeled as a Poisson process \cite{3g}, and hence, the energy consumption of a device can be seen as a semi-regenerative process where the regeneration point is at the end of each successful data transmission.
For node $i$, the remaining energy at time $t$ is denoted by $E_i(t)$, the average payload size by $D_i $, and the power consumption in transmission mode by $\xi P_{i}+P_c$, where $P_c$ is the circuit power consumed by electronic circuits, $\xi$ is the inverse of power amplifier (PA) efficiency, and $P_{i}$ is the transmit power for reliable data transmission. 
We define the {\it expected lifetime} for node $i$ at the regeneration point as the multiplication of reporting period by the ratio between remaining energy and the average energy consumption per reporting period, as follows:
\begin{align}
L_i(t) \buildrel \Delta \over = &\frac{E_i(t)}{\mathcal E_s^i+\mathcal E_d^i}T_i,\label{lif1}
\end{align}
where
$\mathcal E_d^i$ is the average energy consumption per reporting period for data transmission:
$$\mathcal E_d^i=[P_c+\xi P_{i}]{D_i}/{R_i},$$
 $R_i$ is the average data transmission rate, $T_i$ the expected length of one reporting period, and $\mathcal E_s^i$ the average static energy consumption in each reporting period for data gathering, processing, and etc. 
    

\subsection{Network Lifetime Definition}\label{defs}
The network lifetime is the  time between the reference time and when a network is considered to be nonfunctional. The instant at which an M2M network is considered to be nonfunctional is application-specific. For example, in safety-critical applications where losing even one node deteriorates the performance or coverage, or in sparse sensor deployments where correlation between gathered data by different nodes is low, the shortest individual lifetime (SIL) may specify the network lifetime. In other cases, e.g. where correlation between gathered data by different nodes is high, the longest individual lifetime (LIL) or the average individual lifetime (AIL) may be defined as the network lifetime. 
Here, we present our derivations for the case in which, the shortest  individual lifetime is considered as the network lifetime, i.e. $L_\text{net}^\text{sil}(t)=\min_{i} L_i(t);$
however, as we will show in section \ref{sc} and \ref{sim}, our proposed lifetime-aware resource allocation framework can be also used with other network lifetime definitions.


 \section{Scheduling for Narrow-Band M2M Networks }
 To facilitate the understanding of the fundamental dependence of network lifetime on the remaining energy, reporting period, channel gain, circuit power, and bandwidth, here we investigate SIL-aware scheduling  for a narrow-band M2M system. Examples of such systems are 2G GSM-based M2M networks, LTE networks in which a specific carrier is reserved for an MTC application, and proprietary M2M networks. In these systems, as at most one node occupies the whole bandwidth at  each time, the uplink scheduling is equivalent to finding the transmission time for each node. Denote the length of the resource pool in time domain as $\tau$, the bandwidth as $w$, and the allocated fraction of time  for data transmission of node $i$ as $\tau_i$. Then, the lifetime expression for node $i$ is found from  \eqref{lif1}, where $\mathcal E_d^i=\tau_i[P_c+\xi P_{i}].$
Denote the signal-to-interference-plus-noise ratio (SINR) for reliable transmission of $D_i$ bits in $\tau_i$ seconds  to be $\eta_i= S(D_i/\tau_i)$. For example, using Shannon capacity formula the data rate function is derived as:
$$R_i=w\log(1+ \frac{\eta_i}{\Gamma_{\text{mcs}}}),$$
and correspondingly  $S(x)$ is derived as:
\begin{equation}
S(x)=\big[2^{\frac{x}{ w}}-1\big]\Gamma_{\text{mcs}}.\label{seq}
\end{equation}
In this expression, $\Gamma_{\text{mcs}}$ is the SNR gap between the channel capacity and a practical modulation and coding scheme (MCS), as investigated in \cite{guo}. One sees in \eqref{seq} that $S(x)$ is strictly convex in $x$
and $S(0) = 0$. Thus, we do not choose a specific MCS, and hence, do not specify the exact form of $S(x)$ in our analysis. Instead, we only
assume $S(x)$ to be strictly convex in $x$ and
$S(0)=0$. Denote the  channel gain between node $i$ and the BS as $h_i$. Then, the required transmit power for node $i$ will be $$P_i={\eta_i [N_0+I]w}/{[h_iG_{tr}]},$$
where $G_{tr}$ is the multiplication of transmit and receive antenna gains, and the  power spectral densities (PSDs)  of noise and interference  at the receiver are denoted by $N_0$ and $I$, respectively. 
Then, the  scheduling optimization problem that maximizes the network lifetime  is formulated as follows:
\begin{align}
\text{maximize}_{\tau_i} &  \hspace{2mm} L_\text{net}^{sil}(t)\label{op1}\\
  \text{subject to: }
   \text{C}.\ref{op1}\text{.1:}&\hspace{1mm}  \sum\nolimits_{i\in \mathcal A}\tau_i \le \tau,\nonumber\\
\text{C}.\ref{op1}\text{.2:}&\hspace{1mm} \tau_i^{m}\le \tau_i  \quad\forall i\in\mathcal A,\nonumber
\end{align}
where $\tau_i^{m}$ is the minimum required transmission time, and is found as a function of maximum allowed transmit power $P_{\max}$, by solving the following equation: 
$$S(\frac{D_i}{\tau_i^{m}})=\frac{P_{\max} h_i G_{tr}}{[N_0+I]w}.$$
One can define $\mathcal Z$ as an auxiliary variable where $
 \mathcal Z=\max_{i\in{\bf{\mathcal A}}}\hspace{2mm}\frac{1}{L_i(t)},$
 and rewrite \eqref{op1} as:
\begin{align}
\text{minimize}_{\tau_i}& \hspace{2mm} \mathcal Z \label{op2}\\
\text{subject to: }& \text{C}.\ref{op1}\text{.1},  \hspace{1mm} \text{C}.\ref{op1}\text{.2}, \text{and} \quad \frac{1}{L_i(t)}\le \mathcal Z, \quad \forall i\in {\bf{\mathcal A}}\nonumber.
\end{align}
Taking the second derivative of the inverse lifetime expression,
$$\frac{\partial^2 1/L_i}{\partial \tau_i^2}=\frac{\xi [N_0+I]w}{E_i(t) T_i h_i G_{tr}}\frac{D_i^2}{\tau_i^3}\ddot S({D_i}/{\tau_i}),$$
one sees that it is a strictly convex function of $\tau_i$ because $\ddot S (x)>0$, where $\dot f (x)$ and $\ddot f (x)$ show the first and second derivatives of function $f(x)$ respectively. Thus, $\mathcal Z$ is also a strictly convex  function of $\tau_i$ because the point-wise maximum operation preserves convexity \cite{boyd_con}. Then, the scheduling  problem in \eqref{op2} is a convex optimization problem, and can be solved using convex optimization tools, as has been investigated in appendix \ref{ape}. In the special case that $S(x)$ is found from \eqref{seq} and $\Gamma_{\text{mcs}}=1$, the real-valued solution of \eqref{tis_e} is found from appendix \ref{ape} as:
\begin{align}
\tau_i^*= \max\big\{\tau_i^m, \frac{\ln(2){ D_i} }{w+ {\mathcal L}(\frac{1}{\mathrm e}\big[\frac{{[h_i G_{tr}][P_c }+{T_iE_i(t)\mu/\lambda_i }]}{{ }{}{\xi (N_0+I)w} }-1\big])w}\big\} , \label{tis}
\end{align}
where $\mathrm e$ is the Euler's number, and $\mathcal L(x)$  is the LambertW function, i.e. inverse of the 
function $f(x) = x\exp(x)$  \cite{lam}.
 Motivated by the facts that: (i) scheduling is done in the time domain; and (ii) nodes which are more critical from network battery lifetime point of view receive a longer transmission time than the minimum required transmission time in order to decrease their transmission powers; the expression in \eqref{tis} represents priority of nodes in uplink scheduling when  network battery lifetime is to be maximized.  From \eqref{tis}, one sees that the priority of nodes  in lifetime-aware  scheduling: 
\begin{itemize} 
 
\item
increases with $\frac{1}{E_i(t)}$ and $\frac{1}{T_i}$, because $\tau_i^*$ increases when the remaining energy, i.e. $E_i(t)$,   decreases or the packet generation rate, i.e. $1/T_i$, increases.
\item
increases with $\frac{1}{h_i}$, because  $\tau_i^*$ increases for devices deployed far from the BS or compensate a large pathloss, in order to reduce the transmit power and save energy.
\item
increases with $D_i$, because $\tau_i^*$ increases in the size of buffered data to be transmitted.
 
\end{itemize}
In the case that there is no constraint on the amount of available radio resources,  the optimal transmission time is found as:
\begin{align}
\tau_i^*= \max\big\{\tau_i^m,\frac{\ln(2){ D_i} }{w+ {\mathcal L}(\frac{1}{\mathrm e}\big[\frac{{P_c[h_i G_{tr}] }}{{ }{}{\xi (N_0+I)w} }-1\big])w}\}. \nonumber
\end{align}
In this case,  when the circuit power consumption increases, the optimal transmission time decreases, and hence, the transmit power increases. Then, for nodes in which  the circuit power  is so high that is comparable with the  transmit power, it is more energy efficient to transmit data with a higher transmit power in order to finish data transmission in a shorter time interval, and hence, reduce the circuit energy consumption. Also, the scheduler must provide time-domain scheduling priority for these devices to decrease their waiting time before receiving services, which  decreases their energy consumption in the idle listening to the base station.  Based on these preliminary insights to the lifetime-aware scheduling problem, in the next section MTC scheduling in both time and frequency domains for SC-FDMA systems is investigated.
 

\section{MTC Scheduling  over SC-FDMA}\label{sc}
 SC-FDMA  is a favorite multiple access scheme  for energy-limited uplink communications.
Using SC-FDMA  implies that:
(i) only $\mathcal G$ clusters of adjacent subcarriers can be allocated to each node\footnote{This is the contiguity constraint \cite{3gpp13}. In existing LTE-A networks, $\mathcal G=1, 2$ are used. As PAPR and spectral efficiency increase in $\mathcal G$ \cite{rel12}, for MTC applications $\mathcal G=1$ is preferred.}; (ii) the transmit power over all assigned subcarriers to a node must be the same \cite{lco}
; and (iii) subcarriers are grouped into chunks, before being assigned to the nodes \cite{opt}. 
Consider the scheduling problem at time $t$, where a set of nodes, i.e. $\mathcal A$, with cardinally $|\mathcal A|$ are to be scheduled for uplink transmission.  Denote the set and total number of available chunks as $\mathcal C$ and $|\mathcal C|$, where each chunk consists of $M$ adjacent subcarriers with a time duration of $\tau$. 
Here, we investigate time- and frequency-domain scheduling, i.e. if the number of resource elements is not sufficient to schedule $\mathcal A$ at once, a subset of $\mathcal A$ is selected, and then, the available resources are assigned to this subset using a frequency-domain scheduler.
The effective SINR for a SC-FDMA symbol is approximated as the average  SINR over the set of allocated subcarriers because each data symbol is spread  over the whole bandwidth \cite{lco}. Then, the achievable data rate for node $i$ is written as:
  \begin{align}
  R(\mathcal C_i, P_i)=|\mathcal C_i| M S_v(\eta_i),
  \end{align}
where $\mathcal C_i$ represents the set of allocated chunks to node $i$, $|\mathcal C_i|$ the cardinality of $\mathcal C_i$, $\eta_i=\frac{P_i G_{tr}h_i^{e}}{|\mathcal C_i| M}$,  $S_v(x)$ is the inverse of $S(x)$ which is a strictly convex and increasing  function of $x$, and hence, $S_v(x)$ is a strictly concave function of $x$ \cite{boyd_con}. 
Also, $h_i^{e}$ is the effective channel gain-to-interference-plus-noise ratio for node $i$ and is defined as \cite{lco}
$$h_i^e=|\mathcal C_i|/\big[ \sum_{j\in \mathcal C_i}\frac{[N_0+I_j]Mw}{h_{ji}}\big], $$
where $h_{ji}$ is the channel gain\footnote{The channel gain of a user over all subcarriers of one chunk is assumed to be constant in ($t,t+\tau$) \cite{lco}.} of node $i$ over chunk $j$, $I_j$ is the PSD of interference on chunk $j$, and $w$ is the bandwidth of each subcarrier.  Thus, using \eqref{lif1} the expected lifetime of node $i$ at time $t+\tau$ is formulated as a function of $P_{i}$ and $\mathcal C_i$ as follows:
\begin{equation}\label{lind}
{L}_i(t+\tau)=
\frac{E_i(t+\tau)-[1-\theta_i] \mathcal E_d^i}{\mathcal E_s^i+\mathcal E_d^i}T_i,
\end{equation}
where 
\begin{align}
&E_i(t+\tau)=E_i(t)-\tau P_c[1-\theta_i]-\frac{D_i}{R(\mathcal C_i, P_i)}\big[P_c+\xi P_i\big]\theta_i\label{econ},
\end{align}
$\theta_i$ is 1 if node $i$ is scheduled with $|\mathcal C_i|>0$  and 0 otherwise, and $[1-\theta_i]\mathcal E_d^i$ is the expected energy consumption  from $t+\tau$ till the successful transmission, i.e. the regeneration point. Now, one can formulate the scheduling problem as:
\begin{align}
\text{maximize}_{\mathcal C_i, P_i,\theta_i} &  \hspace{2mm} L_\text{net}^{sil} (t+\tau)\label{op7}\\
  \text{subject to: }\text{C}.\ref{op7}\text{.1:} \hspace{1mm} & \sum\nolimits_{i\in \mathcal A}|\mathcal C_i|\le |\mathcal C|,\nonumber\\
  \hspace{1mm}\text{C}.\ref{op7}\text{.2:} \hspace{1mm} & \mathcal C_i\text{:    contiguous}\quad \forall i\in \mathcal A,\nonumber\\
 \hspace{1mm}\text{C}.\ref{op7}\text{.3:} \hspace{1mm}  & \mathcal C_i  \cap \mathcal C_j =\emptyset \quad \forall i,j\in \mathcal A, i\ne j, \nonumber\\
  \hspace{1mm}\text{C}.\ref{op7}\text{.4:} \hspace{1mm} & {\theta_i  D_i}/{R(\mathcal C_i,P_i)}\le \tau \quad \forall i\in \mathcal A,\nonumber\\
  \hspace{1mm}\text{C}.\ref{op7}\text{.5:} \hspace{1mm} & P_i\le P_{\max}\quad \forall i\in \mathcal A,\nonumber\\
  \hspace{1mm}\text{C}.\ref{op7}\text{.6:} \hspace{1mm}  & [1-\theta_i] Q_i = 0 \quad \forall i\in \mathcal A, \nonumber
\end{align} 
where $\emptyset$ is the  empty set, and  $Q_i\in \{0,1\}$ is a binary parameter used for prioritizing traffics from high-priority nodes or traffics with zero remaining delay budgets to be transmitted immediately.   The scheduler gives the highest priority to the traffic from node $i$ if  $Q_i=1$. If $Q_i=0$,  node $i$ will be either scheduled by using the remaining radio resources from scheduling high-priority traffic or will be scheduled in later time slots. In \eqref{op7}, C.\ref{op7}.1 is due to the limited set of available uplink radio resources, C.\ref{op7}.2 is due to the contiguity requirement in SC-FDMA, C.\ref{op7}.3 assures each resource element will be at most allocated to one device, C.\ref{op7}.4-5 assure that the $D_i$ bits of data can be transmitted over the assigned set of contiguous resources with a transmit power lower than the maximum allowed transmit power, and C.\ref{op7}.6 assures that high priority traffic will be scheduled ahead of low-priority traffic. One sees that the optimization problem in \eqref{op7} is not a convex optimization problem due to the contiguity constraint, as discussed in \cite{lco}. 
The straightforward solution for this problem consists in   reformulating the problem as a pure binary-integer program.
From \cite{opt}, we know that the complexity of search over all feasible resource allocations for SC-FDMA systems  is 
\begin{equation}\label{eqc}\sum\nolimits_{i=1}^{|\mathcal A|}\big(_{\hspace{1.5mm}i}^{|\mathcal A|}\big)i!\big(_{\hspace{1.5mm}i-1}^{|\mathcal C|-1}\big).\end{equation}
Thus, the straightforward  approach is clearly
 complex, especially for cellular networks with massive machine-type communications. While shifting complexity from device-side to network-side is  feasible for enabling
 massive IoT connectivity in beyond 4G era \cite{5gera}, and quantum-assisted communication for realizing such receivers  has been introduced in \cite{myth,imre}, here we focus on deriving low-complexity scheduling solutions which are applicable  even in existing cellular infrastructures.
 In the following, we present a low-complexity lifetime-aware scheduling solution. 

\subsection{Low-Complexity Scheduling Solution: The Prerequisites } \label{int-alg1}
 Before presenting our proposed solution, 
  in the sequel a set of prerequisites are investigated. 
\subsubsection{Optimized transmit power for a given scheduling}\label{optp}
For a fixed chunk assignment to  node $i$, i.e. $\mathcal C_i$, one can use the results presented in \cite[section~III]{guon} to prove that the presented energy consumption expression for time interval $[t,t+\tau]$ in \eqref{econ} is a strictly quasiconvex function of $P_i$. Then, one can find the transmit power that minimizes the energy consumption in this interval. Using convex optimization theory, the optimal transmit power is found as the maximum of $P_{\min}$ and the solution to the following equation:
$$ S_v(K P_i)-K[P_c/\xi+P_i]\ddot S_v(K P_i)=0,$$
where $K=\frac{{{G_{tr}}h_i^{e}}}{|\mathcal C_i| M},$ and from  \text{C}.\ref{op7}\text{.4}, one can drive $P_{\min}$ as the solution to:
$$ S_v(P_{\min} K)=D_i/[\tau |\mathcal C_i| M].$$
 As an example, when $S(x)$ is found from \eqref{seq}, the optimal transmit power is: 
\begin{align}
\mathcal P(\mathcal C_i)=\min\bigg\{{P_{\max}},
  \max\big\{P_{\min}, \frac{1}{K}\big[\frac{KP_c/\xi-1}{\mathcal L(\frac{KP_c/\xi-1}{\mathrm e})}-1\big]\bigg\},\label{pin}
\end{align}
and $P_{\min}$ is found as: $P_{\min}= \left[2^{ \frac{ D_i}{\tau M|\mathcal C_i| w}}-1\right]/K.$

\subsubsection{Scheduling metric}\label{met}
Let us consider the scheduling problem at time $t$, where $|\mathcal C|-1$ chunks are already allocated to the nodes, $\mathcal C_i$ shows the set of already assigned resources to node $i$, and one further chunk is available to be allocated to the already scheduled or non-scheduled nodes. Given $\mathcal C_i$ and $\theta_i$, the expected lifetime of node $i$ at $t+\tau$ is derived  from \eqref{lind}-\eqref{econ}, as follows:
\begin{align}
&\mathcal F(E_i(t),\mathcal E_d^i, \mathcal  E_s^i, T_i, \mathcal C_i,\theta_i) \buildrel \Delta \over =L_i(t+\tau)=\label{mfd}\\
&\qquad\frac{E_i(t)\hspace{0.5mm}\text{--}\hspace{0.5mm}\tau P_c[1-\theta_i]-\frac{D_i\big[P_c+\xi \mathcal P(\mathcal C_i)\big]}{R\big(\mathcal C_i,\mathcal P(\mathcal C_i)\big)}\theta_i-[1-\theta_i] \mathcal E_d^i}{{\mathcal E_s^i+\mathcal E_d^i}}T_i.\nonumber
 \end{align}
When the SIL network lifetime is to be maximized, the index of  node with highest priority to be scheduled is found as follows:
\begin{align}
L_{net}^{sil}(t+\tau)|_{\text{opt. scheduling}} &> L_{net}^{sil}(t+\tau)|_{\text{any scheduling}}\nonumber\\
\rightarrow\min_{i\in\mathcal A} \{L_i(t+\tau)+\Delta L_i^{i^*}\} &> \min_{j\in\mathcal A} \{L_j(t+\tau)+\Delta L_j^{j^*}\},\nonumber
\end{align}
where  $i^*$ is the index of selected node using optimal scheduler, $j^*$ the index of selected nodes using any other scheduler,  $\Delta L_{i}^{j}$ the change in the lifetime of node $i$ when the extra chunk is allocated to node $j$, and $\Delta L_{i}^{j}=0$ for $i\ne j$. 
Then, we have:
\begin{align}
&\min\{L_{i^*}+\Delta L_{i^*}^{i^*}, L_{j^*},\min_{i\in\mathcal A\setminus \{i^*,j^*\}  }L_i \}>\nonumber\\
&\qquad\qquad\min\{L_{i^*},L_{j^*}+\Delta L_{j^*}^{j^*}, \min_{j\in\mathcal A\setminus \{i^*,j^*\}  }L_j\},\nonumber\\
&\rightarrow\min\{L_{i^*}+\Delta L_{i^*}^{i^*}, L_{j^*} \}> \min\{L_{i^*},L_{j^*}+\Delta L_{j^*}^{j^*}\}\label{1eq-s10},
\end{align}
where the time index is dropped for the sake of notational continence. One sees that the only choice of $i^*$ that satisfies \eqref{1eq-s10}  for any choice of $j^*$ at time $t$ is
\begin{align}
 i^*(t)=&\arg\min_{i\in\mathcal A}\hspace{1mm} L_i(t+\tau)\nonumber\\
 =&\arg\min_{i\in\mathcal A}\hspace{1mm} \mathcal F(E_i(t),\mathcal  E_d^i,\mathcal  E_s^i, T_i, \mathcal C_i,\theta_i)\label{isil}, 
\end{align}
if it satisfies the lifetime improvement constraint: $\Delta L_{i^*}^{i^*}> 0$, i.e. its battery lifetime can be improved by assigning the new chunk.  If the battery lifetime of $i^*(t)$ cannot be improved, we remove $i^*$ from $\mathcal A$, and repeat the criterion in \eqref{isil} in order to find  node with the shortest lifetime that can improve its lifetime using more chunks. 

Denote  by $V_{\mathcal X}$ the expected battery lifetime vector, where $\mathcal X$ is the 
scheduling criterion, and the $i$th element of $V_{\mathcal X}$ shows the expected battery lifetime of node $i$  under  criterion $\mathcal X$. 

 \begin{definition}\label{de1212}
A feasible scheduling  satisfies  the  max-min fairness  criterion  if no  other  scheduling  has  a lexicographically  greater  sorted lifetime vector, i.e.
$V_{\text{max-min}}^s \hspace{1mm}{\ge}_{\text{lex}} \hspace{1mm}V_{\text{any}}^s, $
where the superscript $s$ shows sorting in non-decreasing order \cite{lexico}. 
\end{definition}
In other words,  this definition means  that if we schedule machine nodes under criterion $\mathcal X$, derive their expected battery lifetimes after  scheduling, and sort the  battery lifetimes in  vector $V_{\mathcal X}$; the resulting lifetime vector from max-min fairness scheduling is lexicographically  greater than  the resulting lifetime vector from any other scheduling criterion.
Then, the  smallest resulting battery lifetime from the max-min fairness criterion will be  as  large  as  possible,  and  the  second-smallest resulting battery  lifetime  will be  as  large  as  possible,  and so on. Comparing the proposed iterative structure in this subsection for prioritizing nodes in lifetime-aware scheduling with the definition \ref{de1212} shows that the proposed scheduling procedure achieves the max-min fairness. This is due to the fact that our scheduler first schedules node with the shortest expected lifetime, then schedules node with the second shortest lifetime, and etc. As a result, the increase in battery lifetime of the selected node in each phase will not be at the cost of decrease in the battery lifetime of another node with already shorter battery lifetime.

\subsection{Lifetime-Aware Scheduling Solution: The Procedure }\label{pro}
The basic idea behind our proposed solution is breaking the scheduling problem into two subproblems: (i) satisfying the minimum resource requirement  for the set of high-priority nodes which must be scheduled at time $t$, called $\mathcal A_d$; and (ii) resource allocation for all nodes based on their impacts on  the network lifetime.  Our proposed solution, presented in Algorithm \ref{a1}, solves the first and second subproblems in step 1 and 2 respectively. The first subproblem is a frequency-domain scheduling problem, where the scheduler allocates contiguous resource elements to the nodes which must be scheduled immediately to satisfy their minimum resource requirements. The second subproblem is a time/frequency scheduling problem. As the minimum resource requirement of $\mathcal A_d$ has been already satisfied in step 1, the scheduler in step 2 selects node with the highest impact on the network lifetime among the scheduled and non-scheduled nodes, and allocates contiguous resource elements to it in order to prolong its battery lifetime, and hence, maximizes the network lifetime. In Algorithm \ref{a1}, we call a resource expansion algorithm named \text{\it ExpAlg}, which is presented in Algorithm \ref{a2}. Given the set  of  available resources, i.e. $\mathcal C$, the set of already allocated resources to the $i$th node, i.e. $\mathcal C_i$, and the maximum allowed number of allocated resource clusters to a node, i.e. $\mathcal G$, this algorithm finds the resource element which satisfies the contiguity constraint, and on which, node $i$ has the best SINR.

\begin{algorithm}[t!]
 Initialization\;
- Define $\mathcal A_d$, where $i\in \mathcal A_d$ if $Q_i=1$\; - Define $\mathcal A_d^c$ as $\mathcal A\setminus \mathcal A_d$\;
- $0\to \theta_i$ and $\emptyset \to \mathcal C_i$, $\forall i\in \mathcal A$ \;
\nl Step 1\;
- $\mathcal A \to \mathcal A_d^t$\;
- \While{$\mathcal A_d^t$ is non-empty}{
- $\arg \min_{j\in \mathcal A_d^t} L_j(t) \to j^*$\;
- $\mathcal A_d^t\setminus j^*\to \mathcal A_d^t$, $ 1\to \theta_i$\;
- \While{$D_{j^*}/R\big(\mathcal C_{j^*},P_{\max})\big)>\tau$}{
- $ExpAlg(\mathcal C,\mathcal C_{j^*},\mathcal G)\to c^*$, $c^*\cup \mathcal C_{j^*}\to \mathcal C_{j^*},\mathcal  C\setminus c^*\to \mathcal C$\;
- If $c^*=\emptyset$, then $0\to \theta_i$, $\mathcal A_d\setminus j^*\to \mathcal A_d$, $\mathcal C_{j^*}\cup \mathcal C\to \mathcal  C$, exit the loop\;
}}
\nl Step 2\;
- $\mathcal A_d\cup\mathcal A_d^c\to \mathcal H$\;
- \While{$ \mathcal C$ and $\mathcal H$ are non-empty}{
- $j^*=\arg \min_{j\in \mathcal H}  \mathcal F(E_i(t),\mathcal  E_d^i,\mathcal  E_s^i, T_i, \mathcal C_i,\theta_i)$\;
- \eIf{$\theta_{j^*}\ne 1$}{
- $1\to \theta_{j^*}$\;
- \While{$D_{j^*}/R\big(\mathcal C_{j^*},P_{\max}\big)>\tau$}{ 
- $ExpAlg(C,\mathcal C_{j^*},\mathcal G)\to c^*$, $c^*\cup \mathcal C_{j^*}\to \mathcal C_{j^*}, \mathcal C\setminus c^*\to \mathcal C$\;
- If $c^*=\emptyset$, then $0\to \theta_i$, $\mathcal C_{j^*}\cup \mathcal C\to \mathcal C$, $\mathcal H\setminus j^*\to \mathcal H$, exit the loop\;}}{
- $\frac{D_i[P_c+\xi \mathcal P \hspace{0.5 mm}(\mathcal C_{j^*})]}{R(\mathcal C_{j^*},\mathcal P \hspace{0.5 mm}(\mathcal C_{j^*}))} \to B$\;
- $ExpAlg(C,\mathcal C_{j^*},\mathcal G)\to c^*$\;
- \eIf{$c^*=\emptyset$}{- $\mathcal H\setminus j^*\to \mathcal H$\;}
{-  $c^*\cup \mathcal C_{j^*}\to \mathcal C_{j^*}, \mathcal C\setminus c^*\to \mathcal C$\; 
- If $\frac{D_i[\xi \mathcal P (\mathcal C_{j^*})+P_c]}{R(\mathcal C_{j^*},\mathcal P (\mathcal C_{j^*}))}> B$, then $\mathcal C_{j^*}\setminus c^*\to \mathcal C_{j^*}, \mathcal C\cup c^*\to \mathcal C$, $\mathcal H\setminus j^*\to \mathcal H$\;}}}
\nl $\mathcal P(\mathcal C_i)\to P_i$, $\forall i\in \mathcal A$\;
\nl \Return $P_i, \theta_i, \mathcal C_i, \forall i\in \mathcal A$
 \caption{SIL-aware scheduling for SC-FDMA }\label{a1}
\end{algorithm}

\begin{algorithm}[t!]
- Inputs: $\mathcal C, \mathcal C_i, \mathcal G$\;
- Number of existing resource clusters in $\mathcal C_i$ $\to$ $n$\;
- \eIf{$n<\mathcal G$}{- $c^*=\arg\max_{m\in {\mathcal C}} \frac{h_{mi}}{N_0+I_m}$\;}{
-                Adjacent resource elements to $\mathcal C_i$ $\to \mathcal C$\;
                - $c^*=\arg\max_{m \in \mathcal C}\frac{h_{mi}}{N_0+I_m}$\;}                
\Return $c^*$
 \caption{Resource expansion algorithm ($ExpAlg$) }\label{a2}
\end{algorithm}

\subsection{Extension to Other Network Lifetime Definitions}\label{othE}
The proposed framework in subsection \ref{pro} can be also used with other network lifetime definitions. When the longest individual lifetime  is considered as the network lifetime, the scheduling solution is found from  a modified version of Algorithm \ref{a1}, in which  the $argmin$ operator is replaced with the $argmax$ operator.  In the sequel, we consider two other network lifetime definitions including: (i) average individual lifetime  defined as:
$$L_{net}^{ail}(t)= [{1}/{|\mathcal A|}]\sum\nolimits_{i\in\mathcal A} L_i(t),$$ and sum of the logarithms of individual lifetimes (SLIL) defined as:
$$L_{net}^{slil}(t)=\sum\nolimits_{i\in\mathcal A} \log(L_i(t)).$$
One sees that AIL-aware scheduling  aims at maximizing the average battery lifetime of machine devices without providing fairness among them, while SLIL-aware scheduling  aims at maximizing the average battery lifetime of machine devices with providing proportional fairness among them.  \cite[chapter~4]{gzor}. 
In order to modify Algorithm \ref{a1} for  AIL- and SLIL-aware scheduling, one needs to derive the respective scheduling metrics, which are investigated in the following.  
\subsubsection{Scheduling metric for AIL}
Following the discussion in subsection \ref{met}, when AIL network lifetime is to be maximized,  index of the node with highest priority in scheduling is found as:
\begin{align}
&L_{net}^{ail}(t+\tau)|_{\text{opt. scheduling}} > L_{net}^{ail}(t+\tau)|_{\text{any scheduling}}\nonumber\\
&\rightarrow\sum_{i\in\mathcal A}  L_i+\Delta L_i^{i^*}  > \sum_{j\in\mathcal A} L_j+\Delta L_j^{j^*},\nonumber\\
&\to L_{i^*}+\Delta L_{i^*}^{i^*}+ L_{j^*}+\sum_{i\in\mathcal A\setminus i^*,j^*  }L_i > 
L_{i^*}+L_{j^*}\nonumber\\
&\hspace{40mm}+\Delta L_{j^*}^{j^*}+ \sum_{j\in\mathcal A\setminus i^*,j^*  }L_j ,\nonumber\\
& \rightarrow  \Delta L_{i^*}^{i^*}  > \Delta L_{j^*}^{j^*} \label{1eq-s1},
\end{align}
where the time index is dropped for  the sake of notational convenience. One sees that the only choice of $i^*$ that satisfies \eqref{1eq-s1}  for any choice of $j^*$ at time $t$ is
\begin{align}
 i^*(t)=&\arg\max_{i\in\mathcal A} \Delta L_i^i,\label{iail}\\
=& \arg\max_{i\in\mathcal A}\hspace{2mm}\mathcal F(E_i(t),\mathcal  E_d^i,\mathcal  E_s^i, T_i, \mathcal C_i^+,1)-\nonumber\\
&\qquad\qquad\qquad\mathcal F(E_i(t),\mathcal  E_d^i,\mathcal  E_s^i, T_i,   \mathcal C_i, \theta_i),\nonumber
\end{align}
where $\mathcal C_i^+$ shows the updated  set of assigned chunks to node $i$, i.e. the set of already assigned chunks plus the new chunk.  Then, for each available chunk we  need to find  the node that its lifetime improvement with the extra chunk is higher than the others.

\subsubsection{Scheduling metric for SLIL}
If SLIL network lifetime is the case,  index of the node with highest priority to be scheduled is found as follows:
\begin{align}
&L_{net}^{slil}(t+\tau)|_{\text{opt. scheduling}} > L_{net}^{slil}(t+\tau)|_{\text{any scheduling}}\nonumber\\
&\rightarrow\sum_{i\in\mathcal A} \log( L_i(t+\tau)\text{+}\Delta L_i^{i^*})  > \sum_{j\in\mathcal A} \log(L_j(t+\tau)+\Delta L_j^{j^*}),\nonumber\\
&\rightarrow\prod_{i\in\mathcal A}  L_i(t+\tau)+\Delta L_i^{i^*}  > \prod_{j\in\mathcal A}  L_j(t+\tau)+\Delta L_j^{j^*},\nonumber\\
&\rightarrow  \frac{L_{i^*}(t+\tau)+\Delta L_{i^*}^{i^*}}{L_{i^*}(t+\tau)}  > \frac{L_{j^*}(t+\tau)+\Delta L_{j^*}^{j^*}}{L_{j^*}(t+\tau)}. \label{1eq-s12}
\end{align}
One sees that the only choice of $i^*$ that satisfies \eqref{1eq-s12}  for any choice of $j^*$ at time $t$ is
\begin{align}
 i^*(t)&=\arg\max_{i\in\mathcal A}  \frac{L_{i^*}(t+\tau)+\Delta L_{i^*}^{i^*}}{L_{i^*}(t+\tau)} \nonumber\\
&= \arg\max_{i\in\mathcal A}     \frac{\mathcal F(E_i(t),\mathcal  E_d^i,\mathcal  E_s^i, T_i, \mathcal C_i^+,1)}{\mathcal F(E_i(t),\mathcal  E_d^i, \mathcal E_s^i, T_i, \mathcal C_i,\theta_i)} \label{islil}.
\end{align}
Comparing \eqref{isil}, \eqref{iail}, and \eqref{islil}, one sees how SIL- and SLIL-aware scheduling provide max-min and proportional fairness among machine nodes, respectively.

\subsection{Performance Analysis} \label{pana}
In the outer loop of the first step of Algorithm \ref{a1}, the scheduler iterates over the set of prioritized nodes. In each iteration, it satisfies the minimum resource requirement of a selected node, and hence, the maximum number of iterations in step 1 will be $|{\mathcal C}_d|\le|\mathcal C|$, where $\mathcal C_d$ is the set of allocated resources to $\mathcal A_d$.  In the second step,  the scheduler iterates over the set of remaining resources, and in each iteration it assigns one resource element to a selected node. Then, the maximum number of iterations in the second step is $|\mathcal C\setminus \mathcal C_d|$, and hence, the complexity order of Algorithm \ref{a1} is  $O( |\mathcal C|)$, which is significantly lower than complexity of optimization problem \eqref{op7}, derived in \eqref{eqc}. This complexity reduction comes at the cost of performance degradation in comparison with the non-relaxed problem in  \eqref{op7}. The gap in performance is due to the  fact that the proposed algorithm 1 works in a sequential manner, i.e. it selects the most energy-critical node, allocates the best resource chunk to it, and continues by allocating \textit{neighbor} resource chunks to it\footnote{due to the contiguity constraint} until it becomes satisfied. One can see that while this sequential structure simplifies the solution, it degrades the performance by not selecting a bunch of neighbor chunks at once. In other words, a chunk on which a tagged node has the best SINR may lead to selection of a suboptimal set of neighbor chunks due to the contiguity constraint in SC-FDMA. However, regarding  the fact that in recent releases of LTE like  LTE-A clusters of chunks can be allocated to the users, i.e. $\mathcal G>1$ is available, the proposed scheduler will not suffer much from the contiguity constraint. On the other hand, the main drawback of this algorithm is seemed to be the need for channel state information (CSI), as the level of consumed energy in  CSI exchange with the BS  is comparable with, or even higher than, the consumed energy in actual data transmission.  We discuss this problem in section \ref{fb}, and tackle it  by presenting  limited-feedback requiring variants of Algorithm 1.  
 
  One must note that in algorithm 1 we have used a binary variable for prioritizing different M2M traffic types. In practice, regarding the diverse set of QoS requirements of different M2M applications, scheduler must be able to handle traffic streams with  multi-level priorities, e.g. alarms, surveillance cameras, temperature sensors, and etc. Multi-level prioritized scheduling has been recently investigated in \cite{mostafa}.  Then, design of a hybrid scheduler combining energy preserving features presented in Algorithm 1 with multi-level prioritized scheduling features presented in \cite{mostafa} makes an interesting research direction for our future research.

\section{Lifetime-Aware MTC Scheduling  with Limited Feedback}\label{fb}
In the previous section, we have presented a lifetime-aware scheduling solution which aims at maximizing the network lifetime. This scheme requires the  CSI of machine nodes as well as other communications characteristics  to be available at the BS. However optimal scheduling based on these 
information sets can improve the network lifetime, it requires machine nodes to send several status packets while the actual amount of useful data to be transmitted in many MTC applications is very limited \cite{ltewang}.  Thus, in this section we present a low-complexity low-feedback frequency-domain scheduling solution to be used with other time-domain schedulers. This scheduler aims at maximizing the network lifetime while  the remaining energy level, reporting period, average static energy consumption, and average pathloss for each device are required at the BS. For M2M applications with limited device mobility, the average pathloss, reporting period, and static energy consumption are semi-constant during the lifetime of a device, and hence, the  BS can save them for future use. Also, the energy consumption of each machine device is expected to be very low, then change in the remaining energy will happen in long time intervals and the BS needs to update it in long time intervals, from days to months. Potential applications of this scheduling solution will be presented in subsection \ref{app}. Denote the subset of devices which are   to be scheduled at time $t$  as $\mathcal A$. From \eqref{lif1}, the expected lifetime of node $i$ at time $t$, which is scheduled with $|\mathcal C_i|>0$ chunks, is formulated  as follows:
\begin{equation}\label{lin}
{L}_i(t)=\frac{E_i(t)T_i }{\mathcal E_s^i+ { D_i[\xi P_{i}+P_c]}/{ R(|\mathcal C_i|, P_i)}},
\end{equation}
where 
 \begin{align}
  R(|\mathcal C_i|, P_i)=|\mathcal C_i| M S_v(\frac{P_i}{|\mathcal C_i| M}\frac{{G_{tr}}}{   \gamma_i [N_0+I]w})\label{rfe},
  \end{align}
 and $\gamma_i$ is the distance-dependent path-loss between node $i$ and the BS.
Then, the  problem in \eqref{op7} is rewritten as:
\begin{align}
\text{maximize}_{|\mathcal C_i|, P_i} &  \hspace{2mm} L_\text{net}^{sil}(t)\label{op8}\\
  \text{subject to: }\text{C}.\ref{op8}\text{.1:} \hspace{1mm} & \sum\nolimits_{i\in \mathcal A}|\mathcal C_i|\le |\mathcal C|,\nonumber\\
  \hspace{1mm}\text{C}.\ref{op8}\text{.2:} \hspace{1mm} & \frac{  D_i}{ R(|\mathcal C_i|, P_i)}\le \tau \quad \forall i\in \mathcal A,\nonumber\\
  \hspace{1mm}\text{C}.\ref{op8}\text{.3:} \hspace{1mm} & P_i\le P_{\max}\quad \forall i\in \mathcal A,\nonumber\\
 \hspace{1mm}\text{C}.\ref{op8}\text{.4:} \hspace{1mm}  & |\mathcal C_i| \in \mathbb{N}_0\nonumber,
\end{align} 
where $\mathbb{N}_0$ is the set of non-negative integers. Using \text{C}.\ref{op8}\text{.2}, the minimum resource requirement of  node $i$, i.e. $|\mathcal C_i|^{\min}$, is found by solving the following equation:
\begin{align}
|\mathcal C_i|^{\min} M S_v(\frac{P_{\max}}{|\mathcal C_i|^{\min} M}\frac{{G_{tr}}}{   \gamma_i [N_0+I]w})=D_i/\tau. \label{cons}
\end{align}
One sees that the optimization problem in \eqref{op8} is not a convex optimization problem because of the $P_i/|\mathcal C_i|$ term in  \eqref{rfe}. To find an efficient solution for this problem, one can pursue a similar approach as in Algorithm \ref{a1}.  The overall solution procedure is presented in Algorithm \ref{a3}.  This algorithm first satisfies the minimum resource requirements of all nodes. Then, if the set of remaining chunks, i.e. $\mathcal C_t^n$, is non-empty, it finds node with the shortest individual lifetime that its lifetime can be improved by assigning more chunks and assigns it one more chunk.  In this algorithm,
$$\mathcal F(E_i(t),E_s^i, T_i, |\mathcal C_i|)=\frac{E_i(t)T_i }{\mathcal E_s^i+ \frac{ D_i[\xi \mathcal P(|\mathcal C_i|)+P_c]}{ R(|\mathcal C_i|, \mathcal P(|\mathcal C_i|))}},$$
where the optimized transmit power for a given number of chunks, i.e. $\mathcal P(|\mathcal C_i|)$, is found from \eqref{pin}.  The outputs of this algorithm are ${\bf y}$ and ${\bf p}$ vectors, where the $i$th entries of  them show  the number of allocated chunks and the transmit power for the $i$th node, respectively.

\begin{algorithm}[t!]
\nl Initialization\;
- Derive $|\mathcal C_i|^{\min}, \forall i\in \mathcal A$, from \eqref{cons}\;
- $|\mathcal C_i|^{\min} \to  {\bf y}(i), \hspace{1mm} \forall i\in \mathcal A$\;
- $\mathcal F(E_i(t),E_s^i, T_i, {\bf y}(i))\to {\bf f}(i), \hspace{1mm} \forall i\in \mathcal A$\;
- $\mathcal A\to \mathcal A_t$\;
\nl \While{$\mathcal C_t^n$}{
- $\arg \min_{i\in \mathcal A}\hspace{1mm}{\bf f}(i)\to { m}$\;
- ${\bf y}(m)+1 \to  x$\;
- \eIf{$\mathcal F(E_m(t),E_s^m, T_m, x)> {\bf f}(m)$}{
- $\mathcal C_t^n-1\to 
\mathcal C_t^n$, $x\to{\bf y}(m)$\;
- $\mathcal F(E_m(t),E_s^m, T_m, {\bf y}(m))\to {\bf f}(m)$\;
}{
- $\mathcal A_t\setminus m\to \mathcal A_t$, and $\infty\to {\bf f}(m)$\;
}
- If $\mathcal A_t$ is empty, then $0\to \mathcal C_t^n$\;}
\nl $\mathcal P({\bf y}(i))\to {\bf p}(i), \forall i\in \mathcal A $\;
\nl \Return ${\bf y}$ and ${\bf p}$\;
 \caption{SIL-aware scheduling with limited feedback}\label{a3}
\end{algorithm}

\subsection{Complexity Analysis and Potential Applications}\label{app}

 Algorithm \ref{a3} works over the set of all nodes in the first step, and over the set of remaining  chunks in the later steps. Thus, its complexity order is $O(|\mathcal C|)$.
 Algorithm \ref{a3} can be used along with the specified time-domain schedulers in \cite{cusm} in order to allocate frequency domain resources to energy-limited nodes and prolong the network lifetime. Another important  application of this low complexity scheduler  consists in uplink scheduling for  {\it time-controlled} M2M communications. The 3GPP and IEEE have defined specific service requirements and features for M2M communications where one of the most important ones is the time-controlled feature \cite{sysreq,m2meee}. Based on this feature, the BS can assign a set of resources, which are repeated in time in regular time intervals, to a node based on its QoS requirements \cite{mas}. For M2M applications that support the time-controlled feature, the derived scheduling solutions using Algorithm \ref{a3} are valid for a long time interval because the communications characteristics of machine nodes, e.g. reporting periods, are semi-constant during their lifetimes. Then, for a group of machine devices with similar delay requirements, one can use the lifetime-aware scheduler in Algorithm \ref{a3}, and assign them a set of persistent uplink transmission grants. The interested reader may refer to \cite{mas}, where persistent resource provisioning for  M2M communications has been introduced.
  
 
\section{MTC Scheduling over LTE networks}\label{lte}
Here, we focus on the air interface of  3GPP LTE Release  13 \cite{3gpp13}. In this standard, radio resources for uplink and downlink transmissions are distributed in both time and frequency domains. In the time domain, data transmissions are structured in frames where each frame consists of 10 subframes each with 1 ms length, while in the frequency domain, the
available bandwidth is divided into a number of subcarriers each with 15 KHz bandwidth.
The minimum allocatable resource element in a frame is a physical resource block pair (PRBP) which consists of 12 subcarriers  spanning over one transmission time interval (TTI) \cite{3gpp13}. Each TTI consists of two slots and includes 12 (or 14) OFDM symbols if long (or short) cyclic prefix is utilized. 
Based on the LTE open-loop power control \cite{3gpp13}, each node determines its uplink transmit power using  downlink pathloss estimation as:
 \begin{align}
PowC(|\mathcal C_i|,\delta_i ) = |\mathcal C_i|P_0 \beta_i \gamma_i[2^{\frac{k_s\text{TBS ($|\mathcal C_i|$, $\delta_i$)}}{|\mathcal C_i|N_sN_{sc}}}-1].\label{pil}
\end{align}
In this expression, the number of assigned PRBPs to node $i$ is denoted by $|\mathcal C_i|$, the estimated downlink pathloss by  $\gamma_i$, the compensation factor by $\beta_i$, the number of symbols in a PRBP by $N_s$, and the number of subcarriers in a PRBP by $N_{sc}$. Also, $k_s$ is usually set to 1.25 and the transport block size (TBS) can be found in Table 7.1.7.2.1-1 of \cite{3gpp13} as a function of $|\mathcal C_i|$ and TBS index. The TBS index, $\delta_i \in\{0,\cdots,33\}$, is a function of modulation and coding scheme as  in Table 8.6.1-1 of \cite{3gpp13}. Based on the LTE specification in \cite{3gpp13}, $P_0$  is set based on the required SNR level at the receiver as:
\[P_0=\beta_i[\text{SNR}_{\text{target}}+P_n]+[1-\beta_i] P_{\max},\]
where $P_n=-209.26$ dB is the noise power in each resource block.  Based on these specifications, one can  rewrite the presented scheduling problems in sections \ref{sc} and \ref{fb} in the context of LTE. Also, one sees that by tuning $\mathcal G$ in Algorithm \ref{a2}, our proposed scheduling solutions can be used for both LTE and LTE-A networks which utilize SC-FDMA and clustered SC-FDMA for uplink transmissions, respectively. Let us  consider the scheduling problem  in section \ref{fb} in the context of LTE. For  node $i$,  the expected battery lifetime is found from \eqref{lif1} as:
\begin{align}
L_i(t)=\frac{E_i(t)T_i}{E_s^i+TTI[P_c+\xi PowC(|\mathcal C_i|,\delta_i ) ]}.\label{lil}
\end{align}
Also, the resource allocation problem in \eqref{op8} reduces to finding the optimal $|\mathcal C_i|$ and $\delta_i$ values, as follows:
\begin{align}
\text{maximize}_{|\mathcal C_i|, \delta_i} &  \hspace{2mm} L_\text{net}^{sil}\label{op10}\\
  \text{s.t.: }\text{C}.\ref{op10}\text{.1:} \hspace{1mm} & \sum\nolimits_{i\in \mathcal A}|\mathcal C_i|\le |\mathcal C|,\nonumber\\
  \hspace{1mm}\text{C}.\ref{op10}\text{.2:} \hspace{1mm} &  \bar D_i \le {{\text{TBS}} (|\mathcal C_i|, \delta_i)}, \quad \forall i\in \mathcal A,\nonumber\\
  \hspace{1mm}\text{C}.\ref{op10}\text{.3:} \hspace{1mm} &PowC(|\mathcal C_i|,\delta_i ) \le P_{\max},\quad \forall i\in \mathcal A,\nonumber\\
 \hspace{1mm}\text{C}.\ref{op10}\text{.4:} \hspace{1mm}& \delta_i\in\{0,\cdots, 33\}; |\mathcal C_i|\in\{1,\cdots,|\mathcal C|\}, \quad \forall i\in \mathcal A,\nonumber
\end{align}
where $|\mathcal C|$ is the total number of available PRBPs, $ \bar D_i={ D_i} +D_{oh}$, and $D_{oh}$ is the size of  overhead information for User Datagram Protocol (UDP), Internet Protocol (IP), and etc.   In order to solve this problem, we can use a modified version of  Algorithm \ref{a3}. The   solution procedure  is presented in Algorithm \ref{a4}. In this algorithm, $\mathcal F(x,y)=L_i(t)\big|_{|\mathcal C_i|=x,\delta_i=y}$, and $|\mathcal C_i|^{\min}$ is the minimum PRBP requirement for node $i$ found as:
\begin{align}
|\mathcal C_i|^{\min}&= \text{minimize}_{\delta_i}\hspace{1mm} |\mathcal C_i|,\label{cminn}\\
\text{subject to: } &\text{TBS}(|\mathcal C_i|,\delta_i)\ge \bar D_i; PowC(|\mathcal C_i|,\delta_i ) \le P_{\max}.  \nonumber
\end{align}
Also, given the assigned number of PRBPs to node $i$, i.e. $|\mathcal C_i|$, and the queued data length as $D_i$,   Algorithm \ref{a4} calls function $\text{\it FunD} (|\mathcal C_i|, D_i)$ in order to derive the corresponding TBS index $\delta_i^*$ that minimizes the transmit power as:
\begin{align}
\delta_i^*\buildrel \Delta \over =  &\text{\it FunD}(|\mathcal C_i|,D_i)=\text{minimize} \hspace{1mm}\delta_i,\label{dminn}\\
&\text{subject to:}\quad \text{TBS}(|\mathcal C_i|,\delta_i)\ge  D_i+D_{oh}.\nonumber
\end{align}  
$\delta_i^*$ can be found by referring to the $(|\mathcal C_i|)$-th column of the TBS table in \cite{3gpp13}, and finding the minimum TBS index for which, the constraint in \eqref{dminn} is satisfied.

 \subsection{Low-Complexity Solution}\label{lowcl}
We can also use linear relaxation in order to transform the discrete optimization problem in \eqref{op10} to a continuous optimization problem. 
Let us introduce an auxiliary variable $\mathcal Z$ and rewrite the optimization problems in \eqref{op10} as follows:
\begin{align}
\text{minimize}_{|\mathcal C_i|} &  \hspace{2mm}  \mathcal Z \label{op11}\\
  \text{subject to: }\text{C}.\ref{op11}\text{.1:} \hspace{1mm} & \sum\nolimits_{i\in \mathcal A}|\mathcal C_i|\le |\mathcal C|,\nonumber\\
 \hspace{1mm}\text{C}.\ref{op11}\text{.2:} \hspace{1mm}&  |\mathcal C_i|^{\min}\le |\mathcal C_i|,   \quad \forall i\in \mathcal A\nonumber,\\
 \hspace{1mm}\text{C}.\ref{op11}\text{.3:} \hspace{1mm}& \mathcal Z\le L_i(t), \quad \forall i\in \mathcal A,\nonumber
\end{align} 
where $\mathcal Z=\max\hspace{1mm} \frac{1}{L_i(t)}.$
From the TBS table in \cite{3gpp13}, one sees that the maximum TBS for one PRBP is 968, then
$$|\mathcal C_i|^{\min}=\max\{ |\mathcal C_i|^{m}, { \bar D_i}/{968}\},$$
in which $|\mathcal C_i|^{m}$ is found by satisfying \text{C}.\ref{op10}\text{.2} and \text{C}.\ref{op10}\text{.3} with equality, as follows:
$$|\mathcal C_i|^{m}P_0 \beta_i \gamma_i[2^{\frac{k_s \bar D_i}{|\mathcal C_i|^{m}N_sN_{sc}}}-1]= P_{\max}.$$
The scheduling problem in \eqref{op11} is a convex optimization problem because the objective function is   concave   and the constraints make a convex set. Then, one can use the dual Lagrangian scheme  and find the desired solution as: 
\begin{align}
|\mathcal C_i|^*= \max\big\{|\mathcal C_i|^{\min},\frac{k_s \ln(2) \bar D_i /[N_sN_{sc}]}{1+\mathcal L\big(\frac{E_i(t)T_i\mu}{\mathrm e P_0\beta \gamma_i\lambda_i \xi TTI}-\frac{1}{\mathrm e }\big)} \big\},\label{cis1}
\end{align}
where $\mu$ and $\lambda_i$:s are Lagrange multipliers.
The derived $|\mathcal C_i|^*$ values  from \eqref{cis1} are fractional solutions to the relaxed problem in \eqref{op11}. Then, we can use randomized rounding  to find the number of assigned PRBPs to each node \cite{round}. Given the assigned number of PRBPs to node $i$ as $|\mathcal C_i|^*$ and the queued data length as $D_i$, the optimal TBS index is found 
from \eqref{dminn}  as $\delta_i^*=\text{\it FunD}(|\mathcal C_i|^*,D_i).$ Then, the corresponding transmit power for node $i$ is computed by inserting the derived $|\mathcal C_i|^*$, $\delta_i^*$, and TBS($|\mathcal C_i|^*,\delta_i^*$) in \eqref{pil}.

 \begin{algorithm}[t!]
\nl Initialization\;
- Derive $|\mathcal C_i|^{\min}, \forall i\in \mathcal A,$ from \eqref{cminn}\;
- $|\mathcal C_i|^{\min} \to  {\bf y}(i), \hspace{1mm} \forall i\in \mathcal A$\;
- {\it FunD}$({\bf y}(i),D_i) \to$ $\delta_i^*$, $\forall i\in \mathcal A$\; 
- $PowC({\bf y}(i),\delta_i^* )    \to {\bf p}{(i)}$,   $\hspace{1mm}\forall i\in \mathcal A$\;
- $\mathcal F  ({\bf y}(i), \delta_i^*)\to {\bf f}(i), \hspace{1mm}\forall i\in \mathcal A$\;
- $\mathcal A\to \mathcal A_t$\;
\nl \While{$\mathcal C_t^n$}{
- $\arg \min_{i\in \mathcal A}\hspace{1mm}{\bf f}(i)\to { m}$\;
- ${\bf y}(m)+1 \to  x$\;
- {\it FunD}$(x,D_m) \to$ $\delta_m^*$\;
- $PowC(x,\delta_m^*)\to P$\;
- \eIf{$P\le P_{\max}$, and $\mathcal F  (x , \delta_m^*) > {\bf f}(m)$}{
- $\mathcal C_t^n-1\to 
\mathcal C_t^n$, $x\to{\bf y}(m),  P\to{\bf p}(m)$\;
- $\mathcal F  (x, \delta_m^*)   \to {\bf f}(m)$\;
}{
- $\mathcal A_t\setminus m\to \mathcal A_t$, and $\infty\to {\bf f}(m)$\;
}
- If $\mathcal A_t$ is empty, then $0\to \mathcal  C_t^n$\;}
\nl \Return ${\bf y}$ and ${\bf p}$\;
 \caption{Scheduling with limited feedback for LTE }\label{a4}
\end{algorithm}

\begin{table}[t!]
\centering \caption{Simulation parameters }\label{par}
\begin{tabular}{p{5 cm}p{3 cm}}\\
\toprule[0.5mm]
Parameter&Value\\
\midrule[0.5mm]
Cell radius&  500 m\\
Path loss model &$128+38\log_{10}(\frac{r}{1000})$ 
\\
PSD of noise& -174 dBm/Hz\\
System bandwidth & 1.4 MHz\\
Transmission time interval, TTI& 1 ms\\
Number of PRBPs in each TTI & 6 \\ 
$k_s, N_s, N_{sc}$& 1.25, 12, 12  \\
TBS index, $\delta_i$& $\{0,\cdots,26\}$\\
Transport block size& Tab. 7.1.7.2.1-1  \cite{3gpp13}\\
Pathloss compensation factor, $\beta_i$&0.92\\
\hline
Number of nodes &18000\\
Data generation at each node& Poisson, rate 1/300\\
Duty cycle, $T_i$& 300 sec, $\forall i\in \mathcal A$\\
Payload+overhead size,  $ \bar D_i$& 600 Bits\\
Circuit power, $P_c$& 7 dBm\\
$\text{SNR}_{\text{target}}$&1 dB\\
Maximum transmit power, $P_{\max}$ &24 dBm\\
Static energy consumption, $\mathcal E_s^i$& 10 $\mu\text{J}$\\
\bottomrule[0.5mm]
\end{tabular}
\end{table}

\section{Performance Evaluation}\label{sim}
In this section, we apply our proposed scheduling algorithms to a 3GPP LTE-A system and provide simulation results to demonstrate lifetime improvements. The testbed for simulations is based on the uplink of a single cell multi-user 3GPP LTE-A network with 1.4 MHz bandwidth \cite{3gpp13}.  The deployment of machine devices and their traffic model follow the  proposed models in \cite[annex~A]{rel12} for smart metering applications, and are reflected in Table \ref{par}. 
Upon having data to transmit, machine nodes send scheduling request on PUCCH to the BS. As per \cite{mas,cusm}, we consider that part of PUSCH radio resources are assigned to M2M communications. 
Here, we assume  the first two radio frames in each second, i.e. 20 subframes each containing 6 PRBPs, have been reserved for uplink transmissions of  machine devices. The BS schedules machine devices and sends back the scheduling grants on PDCCH to let them transmit their packets in the upcoming reserved resources. Six different MTC scheduling schemes that have been implemented  in our simulations are as follows:

\begin{itemize}

\item
{\bf Scheme 1}: This scheme is based on Algorithm \ref{a1}, and aims at maximizing the SIL network lifetime.

\item
{\bf Scheme 2}: This scheme is based on Algorithm \ref{a3} and \ref{a4}, and provides a low-complexity solution with limited feedback requirement. In this scheme, a round robin (RR) scheduler is used for time-domain scheduling, and Algorithm \ref{a4} is used for frequency-domain scheduling.

\item
{\bf Scheme 3}: This scheme is based on Algorithm \ref{a1}, and  aims at maximizing  the LIL network lifetime. 

\item
{\bf Scheme 4}: This scheme consists of two RR schedulers for time- and frequency-domain scheduling, and represents the delay/priority-aware scheduling  schemes in literature \cite{del,lco} when the MTC traffic has no strict delay/priority requirement.
\item
{\bf Scheme 5}: This scheme consists of a channel-aware scheduler for time-domain scheduling, a RR scheduler for frequency domain scheduling, and represents the proposed channel-aware  scheduling schemes in  \cite{del,chan}.

\item
{\bf Scheme 6}: This scheme represents the proposed energy efficient MTC scheduling algorithm in \cite{coex}, where the ratio between the sum data rates  and the transmit power consumptions of all devices is  maximized.
 \end{itemize}
 
One must note that a fair comparison requires us to compare scheme 1, as a lifetime-aware time/frequency-domain scheduling solution against schemes $\{3,4,6\}$ which either benefit from RR scheduling or lifetime-aware scheduling with full CSI. On the other hand, scheme 2 which   benefits from the low-complexity lifetime-aware solution with limited-CSI, can be compared against scheme 4 and 5, which benefit from RR scheduling and channel-aware scheduling with limited CSI, respectively.

 \begin{figure}[t!]
    \centering
       \centering
        \includegraphics[trim={0.3cm 1.6cm 1cm 0.6cm},clip,width=3.5in]{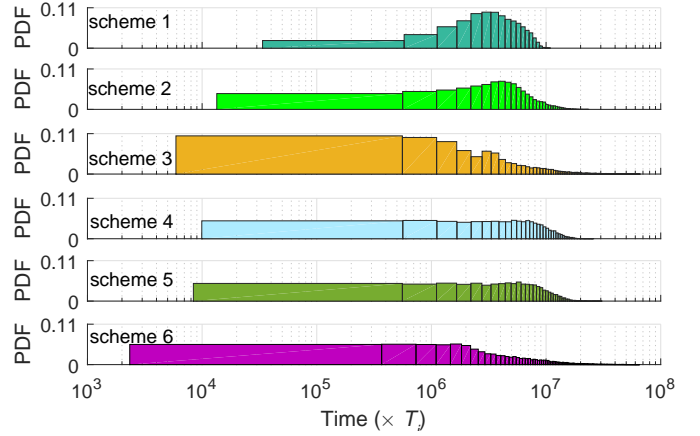}
        \caption{Empirical PDF of individual lifetimes using different scheduling schemes}\label{cdfs}
\end{figure}

 \subsection{Performance Evaluation of the Proposed Schedulers}
 
\begin{figure}[t!]
    \centering
     \includegraphics[trim={.1cm 0cm .1cm 0cm},clip,width=3.1in]{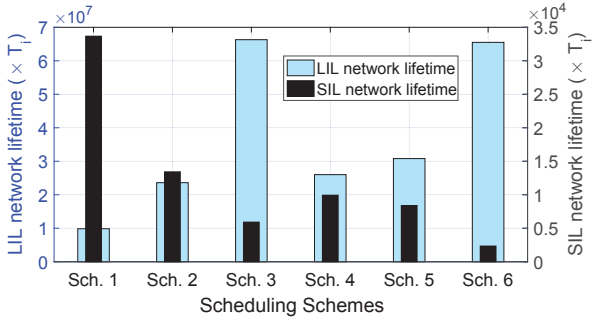}
\caption{Network lifetime comparison  }\label{minx}
    \end{figure}

Fig. \ref{cdfs} represents the probability density function (PDF) of battery lifetimes of machine nodes using different scheduling schemes. The {\it x}-axis has been depicted in log-scale to highlight the differences in PDFs when the initial energy drains happen. One sees that  the first-energy-drain using scheme 1, which aims at maximizing the SIL network lifetime, happens  much later than the first energy drain using the benchmarks, i.e. scheme 4, 5, and 6. Also, one sees that the last energy drain using scheme 3, which aims at maximizing the LIL network lifetime, happens much later than the benchmarks. Furthermore, we see that the PDF of scheme 1 has a compact shape, which shows that the individual lifetimes of machine devices  are distributed in a limited time interval. 
 The detailed SIL network lifetime  comparison of the proposed scheduling schemes is presented in Fig. \ref{minx}.  In this figure, it is evident that the achieved SIL network lifetime from scheme 1 is 2.4 times higher than scheme 4, three times higher than scheme 5, and 13.4 times higher than the scheme 6. Also, one sees that scheme 2, which aims at maximizing the SIL network lifetime with limited feedback requirement, outperforms the baseline schemes 4 and 5. 
 The detailed LIL network lifetime  comparison is presented in Fig. \ref{minx}.  In this figure, it is evident that the achieved LIL network lifetime from scheme 3 is 1.55 times higher than scheme 4, 1.15 times higher than scheme 5, and 0.02 times higher than  scheme 6.  
 
%
%
%

%
 Fairness of the proposed scheduling schemes is investigated in Fig. \ref{varj}. The right axis of Fig. \ref{varj} shows the variance of  individual lifetimes, while
 the right axis  represents the modified Jain's fairness index \cite{jain}, calculated as:
 $$\mathcal J=\frac{(\sum_{i\in \mathcal A} L_i)^2}{|\mathcal{A}|\sum_{i\in \mathcal A} L_i^2}.$$ 
 One sees that scheme 1 achieves the highest fairness index. Recall from  Fig. \ref{minx}, where it was shown that SIL-aware scheduling prolongs the shortest individual lifetime in the network.  Comparing Fig. \ref{minx} and Fig. \ref{varj} indicates that using SIL-aware scheduling, machine nodes will last all together for a long period of time, and will die almost at the same time. 
 
\begin{figure}[t!]
  \centering
     \includegraphics[trim={0cm 0.01cm 0cm 0cm},clip,width=3in]{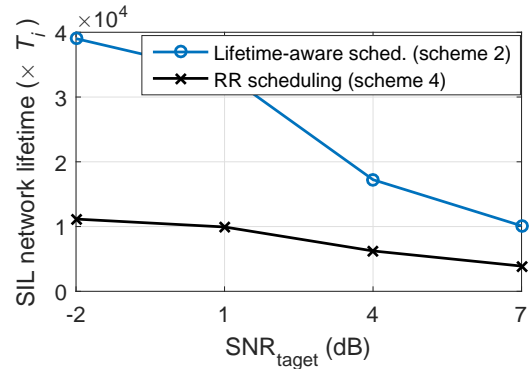}
\caption{Impact of the link budget on lifetime}\label{snr}
    \end{figure}

\begin{figure}[t!]
   \centering
     \includegraphics[trim={0cm 0cm 0cm 0cm},clip,width=3in]{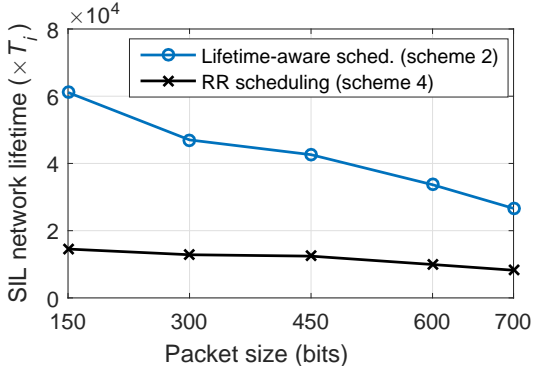}
\caption{Impact of $\bar D_i$ on lifetime}\label{pak}
    \end{figure}

 Fig. \ref{snr} indicates the impact of link budget on the network battery lifetime. Recall the transmit power expression in \eqref{pil}. From this expression, one sees that transmit power is an increasing  function of $\text{SNR}_{\text{target}}$. In Fig. \ref{snr}, one sees that the battery lifetime significantly decreases in $\text{SNR}_{\text{target}}$.  Also, one sees that the achieved battery lifetime from scheme 2 is approximately 2 times higher than the baseline scheme for different $\text{SNR}_{\text{target}}$ values. Similar results can be seen in Fig. \ref{pak} for the impact  of $\bar D_i$ on the network battery lifetime. One sees when  the packet size increases the network  lifetime decreases. Also, one sees that the achieved network lifetime from scheme 2 is  approximately 2 times higher than the baseline scheme for different $\bar D_i$ values.
  
  \subsection{Comparison of Network Lifetime Definitions}\label{compl}
\begin{figure}[!t]
\centering
     \includegraphics[trim={0.8cm 1.2cm 2cm 0cm},clip,width=3.3in]{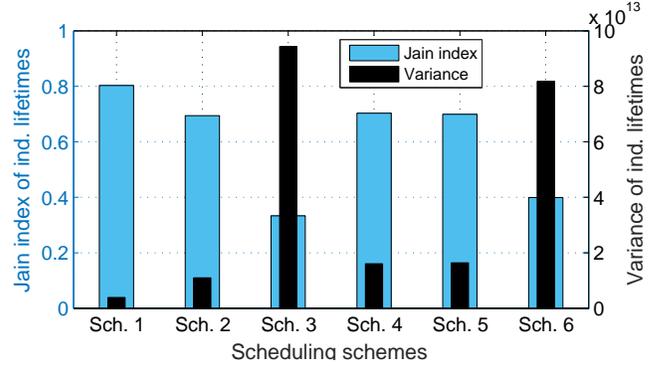}
\caption{Fairness of the proposed schemes}\label{varj}
\end{figure}

 In Fig. \ref{varj}, one sees that by SIL-aware scheduling,  all nodes are expected to have their batteries drained approximately at the same time. This is due to the fact that SIL scheduler  tries to prolong battery lifetimes of low-battery nodes at the cost of sacrificing 
   high-battery nodes' lifetimes. On the other hand, by  LIL-aware scheduling, battery lifetimes of all other nodes are  sacrificed to prolong battery lifetime of the node with the longest battery lifetime. Then, it is clear that depending on the M2M application, the choice of network lifetime to be used for scheduling and consequently the scheduler design, can be different from one network to the other. For example, for a network in which loosing even a small number of nodes deteriorates the performance like sensors installed in urban trash bins, and hence batteries must be replaced when drained,  SIL-aware scheduling may significantly reduce the network maintenance costs by reducing the efforts to monitor the network continuously and replace battery-drained machine nodes one by one. On the other hand, for networks in which correlation between gathered data from different nodes is high like sensors installed in an area for temperature monitoring,   LIL-aware scheduling may minimize the maintenance costs by prolonging battery lifetimes of a subset of nodes.

%
%
%

  \begin{figure}[!t]
        \centering
                \includegraphics[width=3.5in]{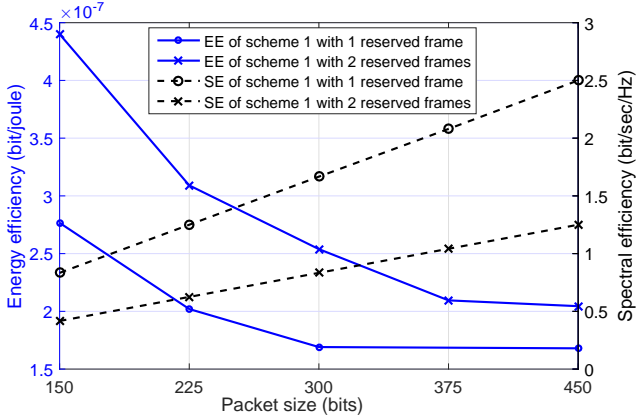}
                \caption{Energy/spectral efficiency tradeoff in MTC resource provisioning}
                \label{tra1}
        \end{figure}
  \begin{figure}[!t]
                \includegraphics[width=3.5in]{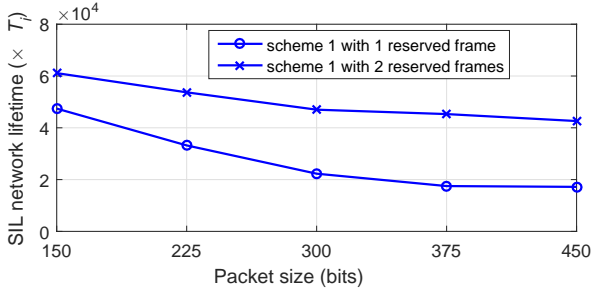}
                \caption{Lifetime tradeoffs in MTC resource provisioning}
                \label{tra2}
\end{figure}
  
 \subsection{Lifetime-Aware Resource Provisioning for MTC}\label{prov}
 As discussed above, provisioning uplink resources for MTC can impact the network lifetime. If the amount of reserved resources is larger than required, some resources are wasted,  spectral efficiency of the network decreases, and the QoS for other services, e.g. web surfing, may be decreased. If the amount of reserved resources is smaller than required, machine nodes must wait for a longer period of time to get access to the reserved resources, and send data with a higher transmit power which in turn reduces their battery lifetimes. Then, one sees that there is  a tradeoff between energy efficiency and spectral efficiency  in uplink transmission. This tradeoff is presented in Fig. \ref{tra1}. In this figure, the solid curves illustrate the energy efficiency of uplink transmissions  in Bit-per-Joule for two resource provisioning approaches: (i) when 1 radio frame consisting of 10 subframes is allocated to MTC in each second; and (ii) when 2 radio frames are allocated to MTC in each second. One sees that the energy efficiency decreases when either the amount of reserved resources decreases or the amount of data to be transmitted over a given set of resources increases. The dashed curves illustrate the  spectral efficiency of uplink transmissions in Bit/Sec/Hz. One sees that spectral efficiency presents a reverse trend when compared with the energy efficiency, i.e. it increases in sending more data in each resource block.  Fig. \ref{tra2} compares the SIL network lifetime. One sees that the network lifetime follows a similar trend to the energy efficiency, i.e. it decreases as the amount of data to be transmitted over a resource block increases. 
%
%
%

\section{Conclusions}
 In this paper, a lifetime-aware resource allocation framework for cellular-based M2M networks is introduced. Theoretical analyses on the impact of scheduling and power control on the energy consumptions of machine nodes, and hence, network lifetime are presented. Based on these analyses, battery lifetime aware  scheduling algorithms are derived. The obtained results show that the optimal scheduling decision depends on the priority class of traffic, remaining battery lifetime of the devices, and the transmission-dependent and -independent sources of energy consumptions. Comparing the lifetime-aware schedulers with the existing scheduling solutions in literature shows that modeling the energy consumption  of MTC, and designing respective scheduling schemes  can significantly prolong the network lifetime. Furthermore, the energy efficiency, spectral efficiency, and network lifetime tradeoffs in uplink MTC scheduling are investigated. It is also shown that uplink scheduling based on the max-min fairness enables machine nodes to last for a long time and  die approximately at the same time, which contributes significantly in network's maintenance costs reduction.   The  results of this article can be used to analyze and optimize the lifetime performance  of deployed machine-type devices over cellular networks.

 \appendices

\section{}\label{ape}
To solve the optimization problem in \eqref{op2}, we first relax C.3.2, solve the relaxed problem, and then apply C.3.2.  The Lagrangian function for the relaxed problem is written as follows:
\begin{equation}\label{tq2}
F=\mathcal Z+\mu[\sum\nolimits_{i\in \mathcal A}\tau_i-\tau]+\sum\nolimits_{i\in \mathcal A}\lambda_i[\frac{1}{L_i(t)}-\mathcal Z],
\end{equation}
where $\mu$ and $\lambda_i$:s are Lagrange multipliers. 
Using convex optimization theory \cite{boyd_con}, the solution for relaxed problem, i.e. $\tau_i^*$, is found by solving:
\begin{align}
&\frac{\partial F}{\partial \tau_i}=0,\quad \to \mu+\lambda_i\frac{\partial L_i^{-1}(t)}{\partial \tau_i}=0,\nonumber\\
 &\to \mu+\frac{\lambda_i}{E_i(t) T_i}\bigg[P_c +\xi[N_0+I]\frac{w}{h_iG_{tr}}S(\frac{D_i}{\tau_i}),\nonumber\\
 &\quad\quad\quad\quad\quad-\frac{D_i}{\tau_i}\xi \dot S(\frac{D_i}{\tau_i})[N_0+I]\frac{w}{h_iG_{tr}}\bigg]=0.\label{tis_e}
\end{align}
 Also, 
 $\mu$ and $\lambda_i$:s, i.e. the Lagrange multipliers,  are found due to the following Karush Kuhn Tucker (KKT) conditions \cite{boyd_con}:
 \begin{align}
&\mu\ge 0;  \hspace{2mm} \big(\sum\nolimits_{i\in \mathcal A}\tau_i-\tau\big)\mu=0;\\
&\big({1}/{L_i(t)}-\mathcal Z\big)\lambda_i=0;  \hspace{2mm} \lambda_i\ge 0; \hspace{2mm} \forall i\in \mathcal A.
  \end{align}
For example, in the special case that $S(x)$ is found from \eqref{seq} and $\Gamma_{\text{mcs}}=1$, the real-valued solution of \eqref{tis_e} is found  as:
\begin{align}
\tau_i^*= \frac{\ln(2){ D_i}/{w}}{1+ {\mathcal L}(\frac{1}{\mathrm e}\big[\frac{{[h_i G_{tr}][P_c }+{T_iE_i(t)\mu/\lambda_i }]}{{ }{}{\xi (N_0+I)w} }-1\big])} , \label{tis}
\end{align}
where $\mathrm e$ is the Euler's number, and $\mathcal L(x)$  is the LambertW function, i.e. inverse of the 
function $f(x) = x\exp(x)$  \cite{lam}.
Now, by applying C.3.2 the optimal transmission time is found as the maximum of $\tau_i^{m}$ and $\tau_i^*$, where $\tau_i^*$ has been introduced in \eqref{op1}.

\ifCLASSOPTIONcaptionsoff
  \newpage
\fi
 
\bibliographystyle{IEEEtran}
\bibliography{IEEEabrv,bibl}

\end{document}